\documentclass[aps,prd,twocolumn,showpacs,floats,floatfix,letterpaper,nofootinbib,superscriptaddress,preprintnumbers]{revtex4}

\usepackage{amssymb,amsmath,latexsym,mathrsfs}
\usepackage{graphicx}
\usepackage{epsfig}
\usepackage{multirow}
\usepackage{array}
\usepackage{color}

\renewcommand{\thefootnote}{\arabic{footnote}}
\begin{document}

\def\thefootnote{\fnsymbol{footnote}}

\title{Primordial Power Spectrum features and $f_{NL}$ constraints}

\author{Stefano Gariazzo}
\affiliation{Department of Physics, University of Torino, Via P. Giuria 1, I--10125 Torino, Italy}
\affiliation{INFN, Sezione di Torino, Via P. Giuria 1, I--10125 Torino, Italy}

\author{Laura Lopez-Honorez}
\affiliation{Theoretische Natuurkunde,\\
Vrije Universiteit Brussel and The International Solvay Institutes,\\
Pleinlaan 2, B-1050 Brussels, Belgium.}

\author{Olga Mena} 
\affiliation{Instituto de F\'isica Corpuscular (IFIC), CSIC-Universitat de Valencia, E-46071, Spain.}

\begin{abstract}

{The simplest models of inflation predict  small non-gaussianities and a featureless power spectrum. However, there exist a large number of well-motivated theoretical scenarios in which large non-gaussianties could be generated. 
In general, in these scenarios the primordial power spectrum will deviate from its standard power law shape. 
We study, in a  model-independent manner, the constraints from future large scale structure surveys on the local non-gaussianity parameter $f_{\rm NL}$ when the standard power law assumption for the primordial power spectrum is relaxed. 
If the analyses are restricted to the large scale-dependent bias induced in the linear matter power spectrum by non-gaussianites, the errors on the $f_{\rm NL}$ parameter could be increased by $60\%$  when exploiting data from the future DESI survey, if dealing with only one possible dark matter tracer. 
In the same context, a nontrivial bias $|\delta f_{\rm NL}| \sim 2.5$ could be induced if future data are fitted to the wrong primordial power spectrum. Combining all the possible DESI objects slightly ameliorates the problem, as the forecasted errors on $f_{\rm NL}$ would be degraded by $40\%$ when relaxing the assumptions concerning the primordial power spectrum shape. 
Also the shift on the non-gaussianity parameter is reduced in this case, $|\delta f_{\rm NL}| \sim 1.6$. 
The addition of Cosmic Microwave Background priors ensure robust future $f_{\rm NL}$ bounds, as the forecasted errors obtained including these measurements are almost independent on the primordial power spectrum features, and $|\delta f_{\rm NL}| \sim 0.2$, close to the standard single-field slow-roll paradigm prediction. }

\end{abstract}
\preprint{IFIC/15-37}

\maketitle

\section{Introduction}
\label{sec:intro}
Inflationary theories have been extremely successful in explaining the horizon problem and the generation of the primordial perturbations seeding the structures of our current
universe~\cite{Guth:1980zm,Linde:1981mu,Starobinsky:1982ee,Hawking:1982cz,Albrecht:1982wi,Mukhanov:1990me,Mukhanov:1981xt,Lucchin:1984yf,Lyth:1998xn,Bassett:2005xm,Baumann:2008bn}. 
The firm confirmation of these theories as the responsible ones for the universe we observe today would come from the detection of a signal of primordial gravitational waves. 
A key observable to disentangle between different inflationary theories is the primordial power spectrum, i.e. the power spectrum of the initial curvature perturbations $P_\mathcal{R}(k)$. 
This power spectrum is usually taken to be a featureless primordial power spectrum (PPS), described by a simple power-law $P_\mathcal{R}(k)\propto k^{n_s-1}$, with $n_s$ the scalar spectral index. 
However, there exists a vast number of models in the literature which may give rise to a  non-standard PPS (see the recent review \cite{Chluba:2015bqa}). 
That is the case of slow-roll induced by phase transitions in the early universe~\cite{Adams:1997de,Hunt:2004vt,Hotchkiss:2009pj}, 
by some inflation potentials~\cite{Starobinsky:1992ts,Leach:2001zf,Gong:2005jr,Adams:2001vc, Chen:2006xjb,Chen:2008wn,Lerner:2008ad,Dvorkin:2009ne,Adshead:2011jq,Hodges:1989dw,Leach:2000yw,Joy:2007na, Jain:2007au,Bean:2008na,
Ashoorioon:2006wc,Ashoorioon:2008qr,
Saito:2008em,Achucarro:2010da,Goswami:2010qu,Brax:2011si,Arroja:2011yu,Liu:2011cw}, 
by resonant particle production~\cite{Chung:1999ve,Mathews:2004vu,Romano:2008rr,Barnaby:2009mc,Barnaby:2010sq},
variation in the sound speed of adiabatic modes~\cite{2013PhRvD..87l1301A,Palma:2014hra} or 
by  trans-Planckian physics~\cite{Brandenberger:2000wr,Danielsson:2002kx,Schalm:2004xg,Greene:2005aj,Easther:2005yr}. 
All these non-standard scenarios, as well as other non-canonical schemes~\cite{Burgess:2002ub,Piao:2003zm,Powell:2006yg,Nicholson:2007by,Lasenby:2003ur,2012JCAP...05..037R,Dvali:2003us,Langlois:2004px}, 
could lead to a PPS which may notably differ from the simple power-law parameterization. 

Another key observable to distinguish among the possible inflationary models is the deviation from the pure Gaussian initial conditions. Non-gaussianities are usually described by a single parameter, $f_{\rm NL}$.  
In the matter dominated universe, the gauge-invariant Bardeen potential on large scales can be parametrized as~\cite{Salopek:1990jq,Gangui:1993tt,Verde:1999ij,Komatsu:2001rj} 
\begin{equation}
\Phi_{\rm NG}=\Phi+f_{\rm NL}\left(\Phi^2-\langle \Phi^2 \rangle\right)\,,
\label{eq:fnl}
\end{equation}
where $\Phi$ is a gaussian random field. The non-gaussianity parameter $f_{\rm NL}$ is often considered to be a constant, yielding non-gaussianities of the {\it local} type. 

Traditionally, the standard observable to constrain non gaussianities is the Cosmic Microwave Background (CMB), through the three point correlation function, or \emph{bispectrum}. 
As the odd power correlation functions vanish for the case of Gaussian random variables, the bispectrum provides the lowest order statistic to test any departure from gaussianity. 
The bispectrum is much richer than the power spectrum, as it depends on both the scale and the shape of the primordial perturbation spectra. The current bound from the complete Planck mission for the local non-gaussianity parameter is $f_{\rm NL}=0.8\pm 5$ ($68\%$~CL)~\cite{Ade:2015ava}. 

The large scale structures of the universe provide an independent tool to test primordial non-gaussianites, as shown in the pioneer works of Refs.~\cite{Dalal:2007cu} and~\cite{Matarrese:2008nc}. 
Dark matter halos will be affected by the presence  of non-gaussianties, and a scale-dependent bias will characterise the non-gaussian signal at large scales~\cite{Slosar:2008hx,Afshordi:2008ru,Carbone:2008iz,Grossi:2009an,Desjacques:2008vf,Pillepich:2008ka,Alvarez:2014vva}. 
The tightest  bounds on primordial non-Gaussianity using exclusively large scale structure data are those obtained from DR8 photometric data, see Ref.~\cite{Leistedt:2014zqa}, which exploits 800000 quasars  and finds $-49 < f_{\rm NL}\ <31$ (see also Ref.~\cite{Agarwal:2013qta}). 
While current large scale structure constraints are highly penalised due to their systematic uncertainties, it has been shown by a number of authors that the prospects from upcoming future large scale structure surveys can reach 
$\sigma(f_{\rm NL})<1$~\cite{Alvarez:2014vva,dePutter:2014lna,Dore:2014cca,Byun:2014cea,Raccanelli:2014awa, Yamauchi:2014ioa,Camera:2014bwa,Ferramacho:2014pua,Ferraro:2014jba,Fedeli:2010ud,Carbone:2010sb,Giannantonio:2011ya}.

Even if Eq.~(\ref{eq:fnl}) is commonly used in the literature assuming a scale independent parameter $f_{\rm NL}$, 
let us mention that some theoretical scenarios can give rise to a scale-dependent $f_{\rm NL}$~\cite{Enqvist:2005pg,Byrnes:2010ft,Riotto:2010nh,Byrnes:2011gh}.
This scale dependence has already been studied in several works, see  e.g.~\cite{LoVerde:2007ri,Biagetti:2013sr,Emami:2015xqa} using large scale structure information, cluster number counts and/or CMB spectral information. 
The forecast on the errors on the non-gaussianity parameters are however known to be parametrization/model dependent~\cite{Biagetti:2013sr,Emami:2015xqa}.
The recent work of Ref.~\cite{Emami:2015xqa} focuses on the complementarity of the different cosmological probes, 
which could help enormously to determine the functional dependence of a scale-dependent non-gaussianity parameter without having to assume a particular choice of such a scale-dependence. 
In particular, they make use of spectral distortions of the CMB background. 
In this work, we shall focus on the forecasts associated to future large scale structure probes only and we will restrict ourselves to a scale independent parameter $f_{\rm NL}$. 
However, when allowing for a non-standard primordial power spectrum as well, additional measurements of the CMB distortion parameters could help in removing some of the degeneracies that appear between non-gaussianities and the parameters governing the primordial power spectrum parameterization. 
Furthermore, these degeneracies could copiously appear in the case of scale-dependent non-gaussinities.
 
Despite the fact that the simplest models of inflation (i.e. single field, slow-rolling with a canonical kinetic term) predict  small non-gaussianities, there are some theoretical scenarios in which large non-gaussianties could be generated, see e.g Ref.~\cite{Bartolo:2004if} and references therein. 
The same deviations from the standard slow-roll inflation that give rise to non-gaussianities could also be a potential source for other features in the PPS~\cite{Hotchkiss:2009pj}, which are absent in the simplest inflation models. 
Particle production during inflation gives rise to both a non-canonical PPS and large non-gaussianities simultaneously~\cite{Barnaby:2010sq}. 
These two phenomena could also appear together in single field models with non-standard inflationary potentials~\cite{Chen:2006xjb,Chen:2008wn,Adshead:2011jq,Saito:2008em,Goswami:2010qu,Arroja:2011yu}, 
as well as in Brane Inflation~\cite{Bean:2008na} and multi-field inflationary models~\cite{Achucarro:2010da}. 
Other possibilities that will give rise to both a non-standard matter power spectrum and non-gaussianities include preheating scenarios~\cite{Chambers:2007se,Bond:2009xx}. 

As nature could have chosen other inflationary scenario rather than the single field slow-roll paradigm, it is interesting to explore, 
in a model-independent way, how the forecasts for large scale structure surveys concerning future measurements of $f_{\textrm{NL}}$ are affected when the assumption of a standard PPS is relaxed. 
This has never been done before while forecasting errors on the $f_{\textrm{NL}}$ parameter and it is a mandatory calculation, because models which will produce non-gaussianities will likely give rise to a non standard PPS  as well. 
Even if non-gaussianties and distortions from the standard power-law PPS are expected to be governed by the same fundamental physics, (and therefore, related to each other),  the underlying inflationary mechanism is unknown a priori. 
A conservative and general approach is therefore to treat these two physical effects as independent and to be determined simultaneously. This is the strategy we follow in this paper. 
The structure of this manuscript is as follows. 
We start describing the parameterization of the  PSS used here  in Sec.~\ref{sec:pps}. Section~\ref{sec:nghb} describes the scale-dependent halo bias in the matter power spectrum, 
while in Section~\ref{sec:method} we describe the methodology followed for our calculations as well as the specifications of the future large scale structure survey illustrated here. We present our results in Sec.~\ref{sec:results} and conclude in Sec.~\ref{sec:concl}.

\section{Primordial power spectrum}
\label{sec:pps}
The simplest models of inflation predict a power-law form for the PPS of scalar and tensor perturbations. 
As previously stated, in principle, a different shape for the PPS (see Ref.~\cite{Chluba:2015bqa} and references therein), can be generated by more complicated inflationary models (see e.g. Ref.~\cite{Martin:2014vha} for some compilation). 
In order to explore the robustness of future forecasted errors from large scale structure surveys on the local non-gaussianity parameter $f_{\textrm{NL}}$, 
we assume a non-parametric form for the PPS, following the prescription of Ref.~\cite{Gariazzo:2014dla}, which is an example of a number of possible methods explored in the literature~\cite{Hunt:2013bha,Hazra:2014jwa,Mukherjee:2003cz, Shafieloo:2006hs,Leach:2005av,Wang:1998gb,Bridle:2003sa,Hannestad:2003zs,Bridges:2008ta,Verde:2008zza,Ichiki:2009zz,Hu:2014aua,2012JCAP...06..006V, Hazra:2013nca,Aslanyan:2014mqa,Matsumiya:2002tx,Kogo:2003yb,Shafieloo:2003gf,TocchiniValentini:2004ht,Nagata:2008tk,Paykari:2014cna,Kogo:2004vt, 2009JCAP...07..011N, 2010JCAP...01..016N,2010JCAP...04..010H,2012JCAP...10..050G,Hazra:2013eva,dePutter:2014hza,Iqbal:2015tta}.
We describe the PPS of the scalar perturbations  by means of a function to interpolate the PPS values in a series of nodes at fixed position. 
The function we exploit to interpolate is commonly named as a \emph{piecewise cubic Hermite interpolating polynomial}, 
the  \texttt{PCHIP} algorithm~\cite{Fritsch:1980}, see the Appendix~A of Ref.~\cite{Gariazzo:2014dla} for details concerning the version of the (\texttt{PCHIP}) algorithm~\cite{Fritsch:1984} used in the following. 
Within this model, one only needs to provide the values of the PPS in a discrete number of nodes and to interpolate among them. As in previous work~\cite{Gariazzo:2014dla}, we define the PPS at twelve nodes, whose values of $k$ are: 
\begin{eqnarray}
k_1     &=& 5 \cdot10^{-6} \  \textrm{Mpc}^{-1} , \nonumber\\
k_2     &=& 10^{-3} \ \textrm{Mpc}^{-1} , \nonumber\\
k_j     &=& k_2 (k_{11}/k_2)^{(j-2)/9} \quad \text{for} \quad j\in[3,10] , \nonumber\\
k_{11}  &=& 0.35 \ \textrm{Mpc}^{-1} , \nonumber\\
k_{12}  &= &10\ \textrm{Mpc}^{-1}~.
\label{eq:nodesspacing}
\end{eqnarray}
In the range $(k_2, k_{11})$, that has been shown to be well constrained by current cosmological data~\cite{dePutter:2014hza},  we choose equally spaced nodes (in logarithmic scale). 
The purpose of the first and the last nodes is to allow for a non-constant behaviour of the PPS outside the well-constrained range. The \texttt{PCHIP} PPS is given by
\begin{equation}
P(k)=P_0 \times \texttt{PCHIP}(k; P_{s,1}, \ldots, P_{s,12})~,
\label{eq:pchip}
\end{equation}
with $P_{s,j}$ the value of the PPS at the node $k_j$ divided by
$P_0=2.2\cdot 10^{-9}$, according to the latest results from the
Planck collaboration, see Ref.~\cite{Ade:2015xua}.

\section{Forecasts}
\label{sec:future}
\subsection{Non-gaussian halo bias} 
\label{sec:nghb}
Non-gaussianities as introduced in Eq.~(\ref{eq:fnl}) induce a scale-dependent bias that affects the matter power spectrum at
large scales. This scale-dependent bias reads
as~\cite{Dalal:2007cu,Slosar:2008hx}
\begin{equation}
\delta_g = b\, \delta_{\rm dm} \quad \mbox{where}\quad b=b_{\rm G}+\Delta b~,
\label{eq:deltaNG}
\end{equation}
where $\delta_g (\delta_{\rm dm})$ are galaxy (dark matter)
overdensities, $b_{\rm G}$ is the gaussian bias and $\Delta b$
reads as
\begin{equation}
\Delta b=3 f_{\rm NL}(1-b_{\rm G})\delta_{\rm c}\frac{H_0^2\Omega_{\rm m}}{k^2 T(k) D(a)}~,
\label{eq:bfnl}
\end{equation}
with $T(k)$ the linear transfer function. The growth factor $D(a)$ is defined as $\delta_{\rm dm}(a)/\delta_{\rm dm}(a=1)$ and $\delta_c$ refers
to the linear overdensity for spherical collapse~\cite{Kitayama:1996pk}.  The power spectrum with non-gaussianties included is obtained using 
\begin{equation}
P_{\textrm{ng}}=P \left(b_G+\Delta b+f\mu_k^2\right)^2~, 
\label{eq:Pstb}
\end{equation}
where $\mu_k$ is the cosine of the angle between the line of sight and
the wave vector $k$ and $f$ is defined as $d\ln
\delta_{\rm dm}/d\ln a$. $P$ is the dark matter power spectrum, whose $k$
dependence is driven either by Eq.~(\ref{eq:pchip}) or by the standard
power-law matter power spectrum (with a given amplitude $A_s$ and slope $n_s$).

\begin{figure}
\begin{tabular}{c}
\includegraphics[width=8.5cm]{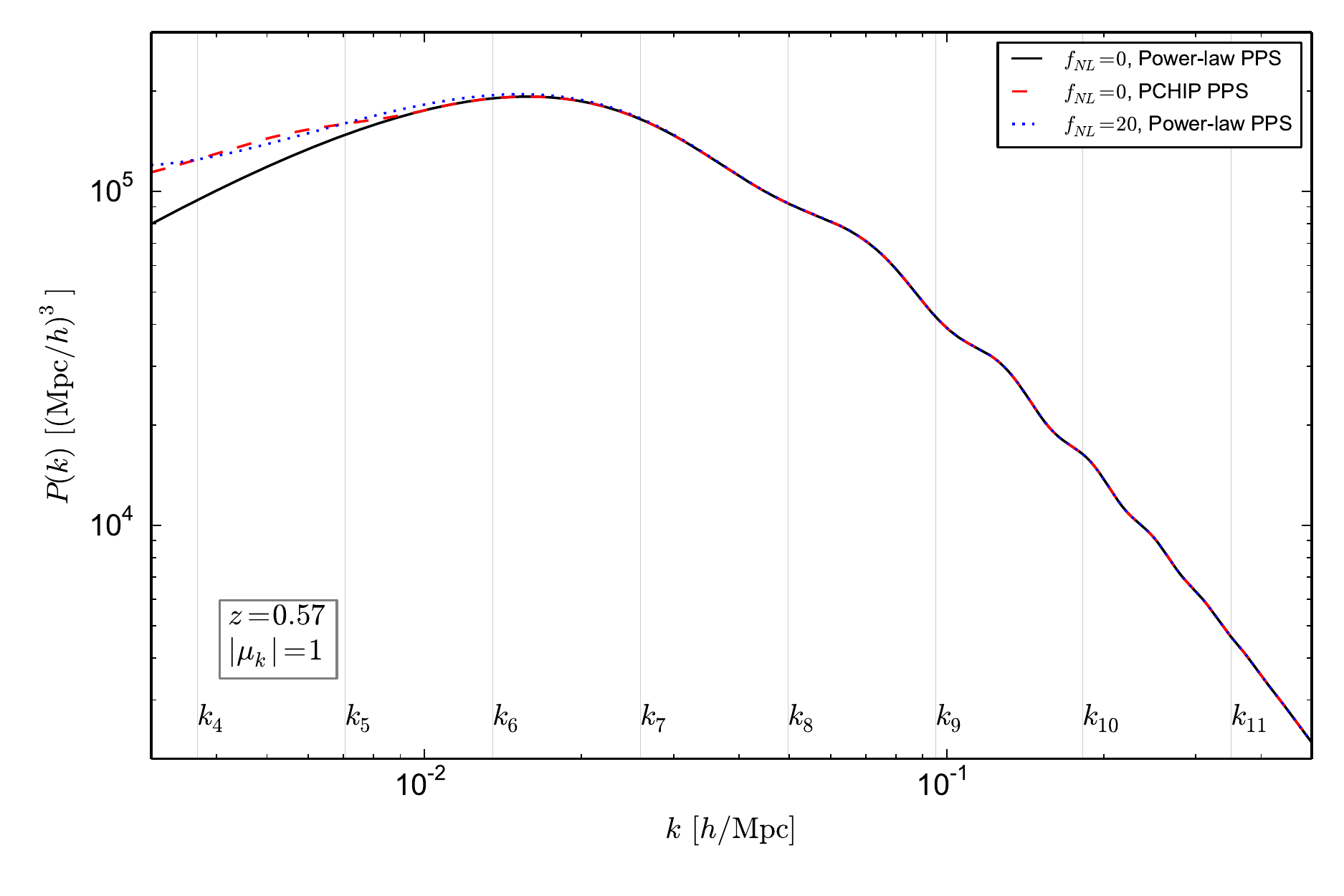}\\
\includegraphics[width=8.5cm]{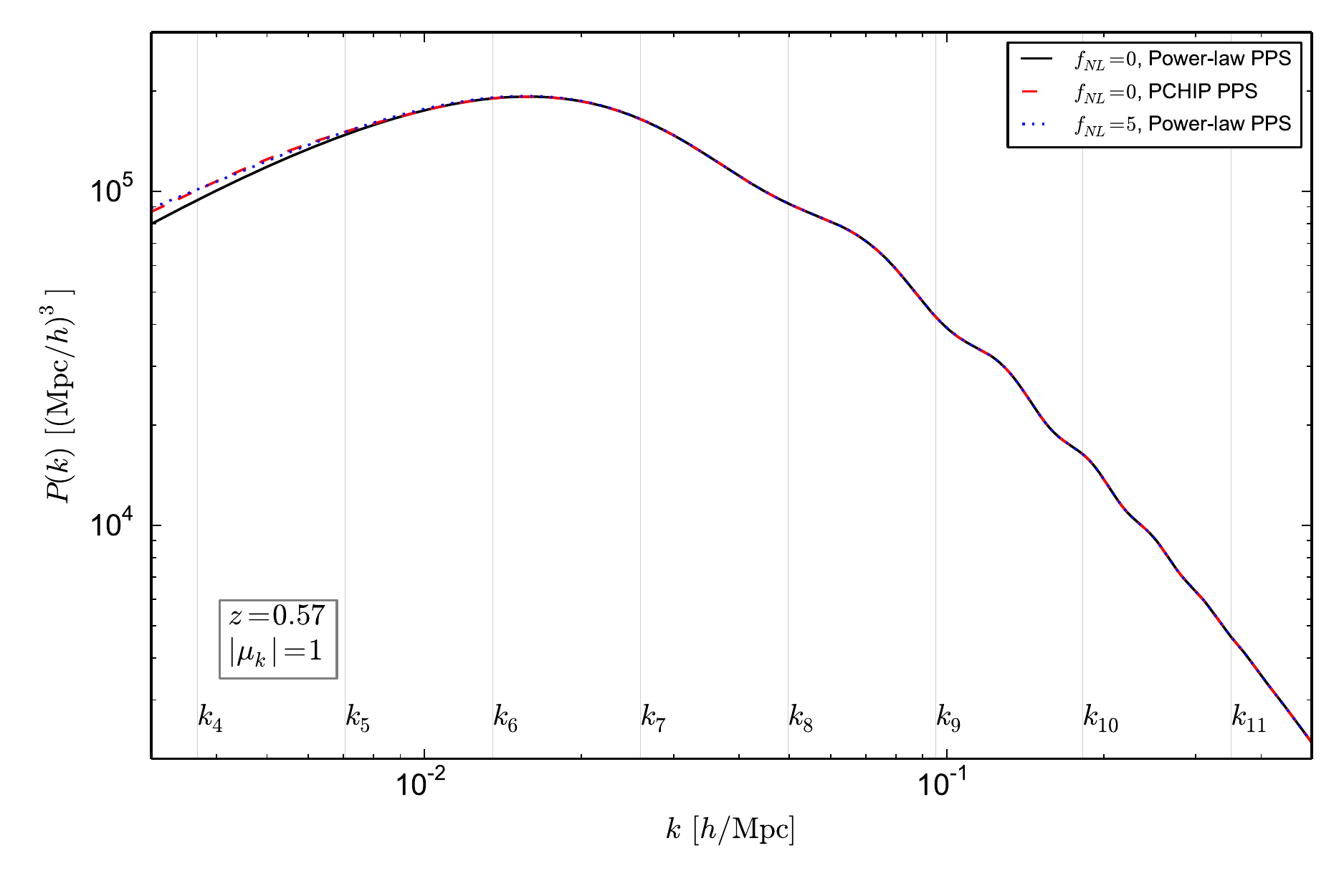}\\
\end{tabular}
\caption{The upper panel depicts the galaxy power spectrum for the standard PPS power law case, for $f_{\rm NL}=0$ (black solid curve) and $f_{\rm NL}=20$ (blue dotted curve), together with a  \texttt{PCHIP} PPS case (red dashed lines) for  $f_{\rm NL}=0$.  
The values of the \texttt{PCHIP} PPS nodes are chosen accordingly to match the predictions of the $f_{\rm NL}=20$ case. 
The lower panel shows the equivalent but for $f_{\rm NL}=5$. We have also changed accordingly the value of the \texttt{PCHIP} PPS nodes. 
We also show with $k_i$ for $i=4,..,11$ the $k$ position of five of the nodes considered in our analysis ($i=5,..,9$), plus $k_{4,10,11}$ that lie outside the $k$ range probed by the DESI experiment. 
The galaxy power spectra are obtained for $z=0.57$,  $|\mu_k|=1$ and assuming a constant gaussian bias $b_G$.}
\label{fig:fig1}
\end{figure}

In Fig.~\ref{fig:fig1}, top panel, we illustrate the galaxy power spectrum in the absence of non-gaussianites (i.e. $f_{\rm NL}=0$) as well as for the $f_{\rm NL}=20$ case. 
We also show, see the thin red dashed line, that using a \texttt{PCHIP} PPS with $f_{\rm NL}=0$ it is possible to match the galaxy power spectrum obtained with a standard power-law PPS and $f_{\rm NL}\neq 0$. 
The $P_{s,j}$ values needed to obtain such an effect were taken to be within their currently $95\%$~CL allowed regions~\cite{Gariazzo:2014dla}. 
Therefore, large degeneracies between the $P_{s,j}$ nodes and $f_{\rm NL}$ parameter are expected. Such a large value of $f_{\rm NL}=20$, albeit allowed by the current large scale structure limits on
local non-gaussianities, is much larger than the expected errors from the upcoming galaxy redshift surveys (see e.g. Refs.~\cite{Byun:2014cea,Sartoris:2015aga}). Therefore, we also illustrate in Fig.~\ref{fig:fig1}, bottom panel, the equivalent plot for $f_{\rm NL}=5$.
For this case, the values for the PPS nodes $P_{s,j}$ required to match the predictions from a standard power law
PPS lie within their $68\%$~CL current allowed regions~\cite{Gariazzo:2014dla}. However, notice that the degeneracies are still present. We therefore expect that the forecasted errors on the $f_{\rm NL}$ parameter are largely affected by the uncertainties on the precise PPS shape.

\subsection{Methodology} 
\label{sec:method}

We focus here on the future spectroscopic galaxy survey DESI (Dark Energy Instrument) experiment~\cite{Levi:2013gra}. 
Although multi-band, full-sky imaging  surveys have been shown to be the optimal setups to constrain non-gaussianities via large scale structure measurements~\cite{Alvarez:2014vva,dePutter:2014lna}, the purpose of the current paper is to explore the degeneracies with the PPS parameterization rather than to optimise the $f_{\rm NL}$ sensitivity. 
Therefore, we restrict ourselves here to the DESI galaxy redshift survey (similar results could be obtained with the ESA Euclid instrument).

In order to compute the expected errors on the local non-gaussianity parameter, we follow here the usual Fisher matrix approach, whose elements, as long as the posterior distribution for the parameters can be well approximated by a Gaussian function, read as~\cite{Tegmark:1996bz,Jungman:1995bz,Fisher:1935bi}
\begin{equation}
\label{eq:fish}
F_{\alpha\beta}=\frac{1}{2}{\rm
Tr}\left[C^{-1}C_{,\alpha}C^{-1}C_{,\beta}\right]~,
\end{equation}
with $C=S+N$ the total covariance. The covariance matrix  contains both the signal $S$ and the noise $N$ terms, and $C_{,\alpha}$ refer to its derivatives with respect to the cosmological parameter $p_\alpha$ in the context of the underlying fiducial cosmology. 
The $68\%$~CL marginalized error on a  given parameter $p_\alpha$  is
$\sigma(p_\alpha)=\sqrt{({F}^{-1})_{\alpha\alpha}}$, with ${F}^{-1}$ the inverse of the Fisher matrix. 
In the following, in order to highlight the differences in the  error on the $f_{\rm NL}$ parameter arising from different PPS choices, we only consider information concerning non-gaussianites from large scale structure data, neglecting the information that could be added from CMB bispectrum measurements.  

Our large scale structure Fisher matrix reads as~\cite{Seo:2003pu}
\begin{eqnarray}  
F^{\rm LSS}_{\alpha \beta}&=&\int_{\vec{k}_{\rm min}} ^ {\vec{k}_{\rm max}} \frac{\partial \ln P_{\rm ng}(\vec{k})}{\partial p_\alpha} \frac{\partial \ln P_{\rm ng}(\vec{k})}{\partial p_\beta} V_{\rm eff}(\vec{k}) \frac{d\vec{k}}{2(2 \pi)^3}  
\label{eq:Fij} \\
&=&\int_{-1}^{1} \int_{k_{\rm min}}^{k_{\rm max}}\frac{\partial \ln P_{\rm ng}(k,\mu_k)}{\partial p_\alpha} \frac{\partial \ln P_{\rm ng}(k,\mu_k)}{\partial p_\beta}  V_{\rm{eff}}(k,\mu_k)\nonumber\\
& &\frac{2\pi k^2 dk d\mu_k}{2(2 \pi)^3}~,  \nonumber 
\end{eqnarray}
where $V_{\rm eff}$ is the effective volume of the survey:
\begin{eqnarray}
V_{\rm{eff}}(k,\mu_k) &=&\left [ \frac{{n}P_{\rm ng}(k,\mu_k)}{{n}P_{\rm ng}(k,\mu_k)+1} \right ]^2 V_{\textrm{survey}},
\label{eq:Veff} 
\end{eqnarray}
where $P_{\rm ng}$ is the power spectrum with non-gaussianities included (see Eq.~(\ref{eq:Pstb})) and $n$ refers to the galaxy number density per redshift bin. We assume $k_{\rm max}=0.1 h$/Mpc and $k_{\rm min}$ is chosen to be equal to $2\pi/V^{1/3}$, where $V$ represents the volume of the redshift bin.
The DESI survey is expected to cover 14000 deg$^2$ of the sky in the redshift range $0.15<z<1.85$, divided in redshift bins of width $\Delta z=0.1$. 
We follow Ref.~\cite{Font-Ribera:2013rwa} for the number densities $n(z)$ and  biases $b_G(z)$ associated to the three types of DESI tracers: Luminous Red Galaxies (LRGs), Emission Line Galaxies (ELGs) and high redshift quasars (QSOs). 
We include the redshift dependence of the (fiducial) bias $b_G$ in Eq.~(\ref{eq:Pstb}) as follows: $b_{G}(z) D(z) = 0.84, 1.7, 1.2$ for ELG, LRG and QSO's respectively, where $D(z)$ is the growth factor as in Eq.~(\ref{eq:bfnl}). 
In order to combine the three different Fishers matrices from the three DESI tracers (LRGs, ELGs and QSOs), we follow the multi-tracer formalism developed in Ref.~\cite{Abramo:2013awa}, where the authors present a generic expression for the Fisher information matrix of surveys with any number of tracers. 
The multi-tracer technique provides constraints that can surpass those set by cosmic variance, due to the differences in the clustering of the possible tracers of large scale structure.

Also, let us remind that the observed size of an object or a
feature at a given redshift $z$  are obtained in terms
of redshift and angular quantities $\Delta z$ and $\Delta \theta$.
These two quantities are related to the comoving dimensions $r_\parallel$ and
$r_\perp$ along and across the line of sight through the angular
diameter distance $D_A(z)$ and the Hubble rate $H(z)$. The same
applies to the Fourier transform associated variables (we will refer to these
as $k_\parallel$ and $k_\perp$ for the dual coordinates of
$r_\parallel$ and $r_\perp$).  Therefore, when one
aims to reconstruct the measurements of galaxy redshifts and
positions in some reference cosmological model which differ from a
given fiducial cosmology, one has to account for geometrical effects
in the following way~\cite{Seo:2003pu} :
\begin{equation}
P_{obs}(k_\parallel^{ref}, k_\perp^{ref})=\frac{D_A(z)|_{ref}^2}{D_A(z)^2}\, \frac{H(z)}{H(z)|_{ref}}\, P_{fid}(k_\parallel, k_\perp)~,
\end{equation}
where the {\it ref} sub/superscript denote quantities in the
reference cosmological model\footnote{$k_\parallel= k_\parallel^{ref}
\,D_A(z)|_{ref}/D_A(z)$ and $k_\perp= k_\perp^{ref} \,
H(z)/H(z)|_{ref}$~.}. We properly account for such effects in our Fisher matrix forecasts when taking numerical
derivatives of the galaxy power spectrum with respect to the
cosmological parameters at given values of $|{\bf k}|$ and $\mu_k$ (or
equivalently, of $k_\parallel$ and $k_\perp$).

In addition to Fisher matrix forecasts, we will also compute the expected shift in the $f_{\rm NL}$ parameter if the $P_{s,j}$ \texttt{PCHIP} parameters (with $j=5,..,9$) are (incorrectly) set to values different from their fiducial ones.  
For that purpose, we use the method developed by the authors of Ref.~\cite{Heavens:2007ka}. 
The main idea is as follows: if the future DESI data are fitted assuming a cosmological model with fixed values of $P_{s,j}$%
\footnote{Fixing the values of $P_{s,j}$ corresponds to fix both $n_s$ and $A_s$ to their best-fit values according to the normalization used here.} 
and therefore characterised by $n^\prime= 5$ parameters ${\cal M}'=\{\Omega_b h^2$,  $\Omega_c h^2$, $h$,  $f_{\rm NL}$, $w\}$, but the true underlying cosmology is a model with different values of $P_{s,j}$ and therefore characterized by $n=10$ parameters 
${\cal M}=\{\Omega_b h^2$,  $\Omega_c h^2$, $h$,  $f_{\rm NL}$, $w$, $P_{s,j}\}$ (with $j=5,..,9$), the inferred values of the $n^\prime=5$ parameters will be shifted from their true values to compensate for the fact that the model used to fit the data is wrong. 
Assuming that the likelihood is gaussian, the shifts in the $n^\prime$  parameters read as~\cite{Heavens:2007ka}
\begin{eqnarray}
\delta\theta'_\alpha =
-(F'^{-1})_{\alpha\beta}G_{\beta\zeta}\delta\psi_\zeta \qquad& &
\alpha,\beta=1\ldots n', \nonumber\\
& & \zeta=n'+1\ldots n \label{offset}~,\qquad
\end{eqnarray}
where $F^\prime$ is the Fisher matrix for the $n'$ parameters model (with the $P_{s,j} $ fixed) and $G$ denotes the Fisher matrix for the $n$ parameters model  (including the $n'$ previous parameters plus the \texttt{PCHIP}  $P_{s,j}$ parameters). 

In the following, unless otherwise stated, we shall adopt the best-fit values from the complete Planck mission~\cite{Ade:2015xua}, which, in the standard power law PPS, corresponds to $A_s=2.2\cdot 10^{-9}$ and $n_s=0.965$ at $k_{pivot}=0.05$. 
Within the \texttt{PCHIP} parameterization, the best-fit values used for the nodes considered in the numerical analyses below  are: $P_{s,5}=1.07099$, $P_{s,6}=1.04687$, $P_{s,7}=1.02329$, $P_{s,8}=1.00024$ and $P_{s,9}=0.97771$. 
These values are obtained calculating the value of the best-fit power-law power spectrum, 
given by Planck 2015 best-fit values for $A_s$ and $n_s$ as mentioned above,
at the positions of the nodes $k_5$ to $k_9$.
The remaining nodes are outside the $k$ range expected to be covered by the DESI survey, given the values of $k_{\rm max}$ and $k_{\rm min}$ considered here.

\subsection{Results} 
\label{sec:results}

In the following, we shall present the results arising from our Fisher matrix calculations, for the two fiducial cosmologies explored here: one in which the PPS is described by its standard power-law form, and a second one in which the PPS is described by the \texttt{PCHIP} parameterization. 
The parameters describing the model with a power-law PPS are the baryon and cold dark matter physical energy densities, $\Omega_b h^2$ and  $\Omega_c h^2$, the Hubble parameter $h$ (with $H_0$, the Hubble constant, defined as $100h$~km/s/Mpc), the scalar spectral index $n_s$, the amplitude of the PPS $A_s$, and the equation of state of the dark energy component $w$. 
The  \texttt{PCHIP} PPS case is also described  by $\Omega_b h^2$, $\Omega_c h^2$, $h$, $w$ plus five nodes $P_{s,j}$ (with $j$ ranging from $5$ to $9$) describing the \texttt{PCHIP} PPS.  
Non-gaussianites of the local type are implemented in both fiducial cosmologies via the  $f_{\rm NL}$ parameter. 
All the results described below (unless otherwise stated) refer to the analysis of the three DESI tracers (ELGs, LRGs and QSOs), i.e. they have been obtained exploiting exclusively the scale-dependent biases imprinted in the power spectra of these three types of tracers. 

\begin{table*}[t]
\begin{tabular}{c|c|c c c c }
\hline
&fiducial & LRG  & ELG & QSO& All   \\
\hline
$\Omega_b h^2$ & $ 0.02267	$ & $ 4.78\cdot 10^{-3}$ & $4.86\cdot 10^{-3} $ & $ 5.11\cdot 10^{-3} $ & $ 2.38\cdot 10^{-3}$ \\
$\Omega_c h^2$ & $0.1131$ & $ 1.75\cdot 10^{-2}$ & $1.65\cdot 10^{-2} $ & $1.51\cdot 10^{-2} $ & $ 7.70\cdot 10^{-3}$ \\
$h	$ & $	0.705$ & $ 5.02\cdot 10^{-2}$ & $ 5.01\cdot 10^{-2}$ & $ 4.69\cdot 10^{-2}$ & $ 2.42\cdot 10^{-2}$ \\
$n_s	$ & $	0.96	$  & $ 5.68\cdot 10^{-2} $ & $4.28\cdot 10^{-2} $ & $ 4.12\cdot 10^{-2}$ & $1.96\cdot 10^{-2} $ \\
$A_s	$ & $	2.2\cdot 10^{-9}$ & $0.341 $ & $0.331 $ & $ 0.302$ & $ 0.156$ \\
$f_{\rm NL}$ & $20$ & $ 19.9$ & $10.1$ & $8.56 $ & $ 4.79$ \\
$w	$ & $	-1	$ & $	5.38\cdot 10^{-2}  $  & $4.09\cdot 10^{-2} $ & $ 6.18\cdot 10^{-2} $ & $2.36\cdot 10^{-2} $ \\
\hline
\end{tabular}
\caption{Marginalized 1-$\sigma$ constraints on  the parameters associated to the standard PPS assuming a fiducial value $f_{\rm NL}=20$. The error on the amplitude of the power spectrum is evaluated on $A_s/(2.2\cdot10^{-9})$.
}
\label{tab:sk1}
\end{table*}

\begin{table*}[t]
\begin{tabular}{c|c|c c c c}
\hline
&fiducial & LRG & ELG & QSO& All \\
\hline
$\Omega_b h^2$ & $ 0.02267$ & $7.85\cdot 10^{-3}$ & $ 3.65\cdot 10^{-3}$ & $4.70\cdot 10^{-3} $& $2.30\cdot 10^{-3}  $\\ 
$\Omega_c h^2$ & $0.1131$ & $2.30\cdot 10^{-2}$ & $1.11\cdot 10^{-2} $ & $ 1.41\cdot 10^{-2} $& $6.36\cdot 10^{-3}  $\\ 
$h	$ & $	0.705$ & $7.67\cdot 10^{-2}$ & $3.59\cdot 10^{-2} $ & $4.62\cdot 10^{-2}  $& $ 2.12\cdot 10^{-2}$\\ 
$P_{s,5}	$ & $	1.07099$ & $ 0.340$ & $0.169$ & $ 0.212$& $ 0.111$\\ 
$P_{s,6}	$ & $	1.04687$ & $ 0.419$ & $ 0.198$ & $0.254$& $0.119$\\ 
$P_{s,7}	$ & $1.02329$ & $ 0.451$ & $0.216$ & $ 0.276$& $ 0.125$\\ 
$P_{s,8}	$ & $	1.00024$ & $ 0.479$ & $0.229 $ & $0.293$& $0.132$\\ 
$P_{s,9}	$ & $	0.97771$ & $ 0.482$ & $0.234$ & $ 0.298$& $ 0.134 $\\ 
$f_{\rm NL}$ & $20$ & $32.2$ & $13.3 $ & $ 12.6$& $ 6.43$\\ 
$w	$ &$ 	-1$ & $4.03\cdot 10^{-2}$ & $2.80\cdot 10^{-2} $ & $4.45\cdot 10^{-2} $& $ 2.45\cdot 10^{-2}$\\ 

\hline
\end{tabular}
\caption{Marginalized 1-$\sigma$ constraints on  the parameters associated to the non-standard PPS assuming  $f_{\rm NL}=20$.}
\label{tab:nsk1}
\end{table*}

\begin{table*}[t]
\begin{tabular}{c|c|c c c c }
\hline
&fiducial & LRG  & ELG & QSO& All   \\
\hline
$\Omega_b h^2$ & $ 0.02267	$ & $2.67\cdot 10^{-4} $ & $2.63\cdot 10^{-4} $ & $ 2.66\cdot 10^{-4}$ & $2.59\cdot 10^{-4}$ \\
$\Omega_c h^2$ & $0.1131$ & $ 1.64\cdot 10^{-3}$ & $1.44\cdot 10^{-3} $ & $1.52\cdot 10^{-3} $ & $ 1.24\cdot 10^{-3}$ \\
$h$ & $0.705$ & $ 6.66 \cdot 10^{-3}$ & $5.24 \cdot 10^{-3} $ & $5.86 \cdot 10^{-3} $ & $ 4.12 \cdot 10^{-3}$\\
$n_s	$ & $	0.96	$  & $6.72\cdot 10^{-2} $ & $ 6.41\cdot 10^{-2}$ & $6.53\cdot 10^{-2} $ & $5.84\cdot 10^{-3}$ \\
$A_s	$ & $	2.2\cdot 10^{-9}	$ & $3.87\cdot 10^{-2} $ & $ 3.28\cdot 10^{-2} $ & $3.51\cdot 10^{-2} $ & $2.71\cdot 10^{-2} $ \\
$f_{\rm NL}$ & $20$ & $ 17.4$ & $9.14 $ & $ 7.58$ & $ 4.56$ \\
$w	$ & $	-1	$ & $	 4.51\cdot 10^{-2} $ &  $ 3.36\cdot 10^{-2}$ & $ 5.44\cdot 10^{-2}$ & $ 2.17\cdot 10^{-2} $ \\
\hline
\end{tabular}
\caption{As Tab.~\ref{tab:sk1} but including CMB priors, see the text for details.}
\label{tab:sk1cmb}
\end{table*}

\begin{table*}[t]
\begin{tabular}{c|c|c c c c c}
\hline
&fiducial & LRG  & ELG& QSO& all \\
\hline

$\Omega_b h^2$ & $ 0.02267$ & $3.92\cdot 10^{-4}$ & $ 3.79\cdot 10^{-4}$ & $3.87\cdot 10^{-4} $& $3.74\cdot 10^{-4} $\\ 
$\Omega_c h^2$ & $0.1131$ & $1.36\cdot 10^{-3}$ & $1.10\cdot 10^{-3} $ & $1.18\cdot 10^{-3} $& $ 1.04\cdot 10^{-3} $\\ 
$h	$ & $	0.705$ & $ 4.13\cdot 10^{-3}$ & $3.14\cdot 10^{-3} $ & $3.62\cdot 10^{-3}  $& $2.93\cdot 10^{-3}  $\\ 
$P_{s,5}	$ & $	1.07099$ & $2.98\cdot 10^{-2}$ & $2.69\cdot 10^{-2} $ & $2.77\cdot 10^{-2}  $& $ 2.60\cdot 10^{-2}$\\ 
$P_{s,6}	$ & $	1.04687$ & $2.89\cdot 10^{-2}$ & $2.10\cdot 10^{-2}  $ & $2.32\cdot 10^{-2}  $& $ 1.99\cdot 10^{-2}$\\ 
$P_{s,7}	$ & $1.02329$ & $2.00\cdot 10^{-2}$ & $1.73\cdot 10^{-2}  $ & $1.84\cdot 10^{-2}  $& $1.69\cdot 10^{-2} $\\ 
$P_{s,8}	$ & $	1.00024$ & $1.92\cdot 10^{-2}$ & $1.76\cdot 10^{-2}  $ & $ 1.86\cdot 10^{-2} $& $ 1.73\cdot 10^{-2}$\\ 
$P_{s,9}	$ & $	0.97771$ & $2.59\cdot 10^{-2}$ & $2.31\cdot 10^{-2}  $ & $ 2.42\cdot 10^{-2} $& $ 2.22\cdot 10^{-2}$\\ 
$f_{\rm NL}$ & $20$ & $13.0$ & $ 6.85$ & $ 5.64$& $4.75 $\\ 
$w	$ &$ 	-1$ & $3.24\cdot 10^{-2}$ & $ 2.46\cdot 10^{-2}$ & $4.0\cdot 10^{-2} $& $2.28\cdot 10^{-2}  $\\ 

\hline
\end{tabular}
\caption{As Table \ref{tab:nsk1} but including CMB priors.}
\label{tab:nsk1cmb}
\end{table*}


\begin{table*}[t]
\begin{tabular}{c|c|c c c c }
\hline
&fiducial & LRG  & ELG & QSO& All   \\
\hline
$\Omega_b h^2$ & $ 0.02267	$ & $4.78\cdot 10^{-3} $ & $ 5.17\cdot 10^{-3} $ & $ 5.18\cdot 10^{-3}$ & $ 2.45\cdot 10^{-3}$ \\
$\Omega_c h^2$ & $0.1131$ & $1.73\cdot 10^{-2} $ & $1.73\cdot 10^{-2}  $ & $1.52\cdot 10^{-2}  $ & $ 7.88\cdot 10^{-3}$ \\
$h	$ & $	0.705$ & $ 5.0\cdot 10^{-2}$ & $ 5.29\cdot 10^{-2}$ & $ 4.75\cdot 10^{-2}$ & $ 2.48\cdot 10^{-2}$ \\
$n_s	$ & $	0.96	$  & $ 5.59\cdot 10^{-2} $ & $ 4.40\cdot 10^{-2} $ & $ 4.11\cdot 10^{-2} $ & $2.0\cdot 10^{-2} $ \\
$A_s	$ & $	2.2\cdot 10^{-9}$ & $0.339 $ & $ 0.347$ & $ 0.305$ & $ 0.160$ \\
$f_{\rm NL}$ & $5$ & $18.9 $ & $9.32 $ & $ 7.83$ & $ 4.45$ \\
$w	$ & $	-1$ & $	5.38\cdot 10^{-2} $  & $ 4.13\cdot 10^{-2}$ & $ 6.19\cdot 10^{-2}$ & $2.38\cdot 10^{-2} $ \\
\hline
\end{tabular}
\caption{Marginalized 1-$\sigma$ constraints on  the parameters associated to the standard PPS assuming a fiducial value $f_{\rm NL}=5$. The error on the amplitude of the power spectrum is evaluated on $A_s/(2.2\cdot10^{-9})$.}
\label{tab:sk5}
\end{table*}

\begin{table*}[t]
\begin{tabular}{c|c|c c c c}
\hline
&fiducial & LRG  & ELG& QSO& All \\
\hline
$\Omega_b h^2$ & $ 0.02267$ & $7.72\cdot 10^{-3}$ & $ 3.61\cdot 10^{-3}$ & $4.61\cdot 10^{-3} $& $2.31\cdot 10^{-3}  $\\ 
$\Omega_c h^2$ & $0.1131$ & $2.28\cdot 10^{-2}$ & $1.09\cdot 10^{-2} $ & $ 1.38\cdot 10^{-2} $& $6.37\cdot 10^{-3}  $\\ 
$h	$ & $	0.705$ & $7.56\cdot 10^{-2}$ & $3.54\cdot 10^{-2} $ & $4.52\cdot 10^{-2}  $& $ 2.13\cdot 10^{-2}$\\ 
$P_{s,5}	$ & $	1.07099$ & $ 0.342$ & $0.169$ & $ 0.215$& $ 0.113$\\ 
$P_{s,6}	$ & $	1.04687$ & $ 0.415$ & $ 0.196$ & $0.251$& $0.120$\\ 
$P_{s,7}	$ & $1.02329$ & $ 0.445$ & $0.212$ & $ 0.270$& $ 0.126$\\ 
$P_{s,8}	$ & $	1.00024$ & $ 0.472$ & $0.225 $ & $0.287$& $0.133$\\ 
$P_{s,9}	$ & $	0.97771$ & $ 0.476$ & $0.230$ & $ 0.292$& $ 0.135 $\\ 
$f_{\rm NL}$ & $5$ & $29.3$ & $11.9$ & $ 10.7$& $ 5.97$\\ 
$w	$ &$ 	-1$ & $4.02\cdot 10^{-2}$ & $2.79\cdot 10^{-2} $ & $4.45\cdot 10^{-2} $& $ 2.44\cdot 10^{-2}$\\ 

\hline
\end{tabular}
\caption{Marginalized 1-$\sigma$ constraints on  the parameters associated to the non-standard PPS assuming  $f_{\rm NL}=5$.}
\label{tab:nsk5}
\end{table*}

\begin{table*}[t]
\begin{tabular}{c|c|c c c c }
\hline
&fiducial & LRG & ELG & QSO& All   \\
\hline
$\Omega_b h^2$ & $ 0.02267	$ & $2.67\cdot 10^{-4} $ & $ 2.63\cdot 10^{-4}$ & $ 2.67\cdot 10^{-4}$ & $2.59\cdot 10^{-4}$ \\
$\Omega_c h^2$ & $0.1131$ & $1.64\cdot 10^{-3} $ & $1.43\cdot 10^{-3} $ & $1.52\cdot 10^{-3} $ & $ 1.24\cdot 10^{-3}$ \\
$h$ & $0.705$ & $ 6.66 \cdot 10^{-3} $ & $ 5.23 \cdot 10^{-3}$ & $  5.85 \cdot 10^{-3}$ & $ 4.11 \cdot 10^{-3}$\\
$n_s	$ & $	0.96	$  & $ 6.71\cdot 10^{-3}$ & $6.40\cdot 10^{-3} $ & $6.53\cdot 10^{-3}  $ & $5.84\cdot 10^{-3}$ \\
$A_s	$ & $	2.2\cdot 10^{-9}	$ & $ 3.87\cdot 10^{-2} $ & $ 3.27\cdot 10^{-2}$ & $3.51\cdot 10^{-2} $ & $2.70\cdot 10^{-2} $ \\
$f_{\rm NL}$ & $5$ & $ 16.8$ & $8.56 $ & $ 7.12$ & $ 4.27$ \\
$w	$ & $	-1	$ & $	4.50\cdot 10^{-2} $ &  $3.36\cdot 10^{-2} $ & $ 5.43\cdot 10^{-2} $ & $ 2.17\cdot 10^{-2} $ \\
\hline
\end{tabular}
\caption{As Tab.~\ref{tab:sk5} but including CMB priors, see the text for details.}
\label{tab:sk5cmb}
\end{table*}

\begin{table*}[t]
\begin{tabular}{c|c|c c c c c}
\hline
&fiducial & LRG  & ELG& QSO& all \\
\hline

$\Omega_b h^2$ & $ 0.02267$ & $3.92\cdot 10^{-4}$ & $ 3.79\cdot 10^{-4}$ & $3.86\cdot 10^{-4} $& $3.75\cdot 10^{-4} $\\ 
$\Omega_c h^2$ & $0.1131$ & $1.36\cdot 10^{-3}$ & $1.10\cdot 10^{-3} $ & $1.18\cdot 10^{-3} $& $ 1.04\cdot 10^{-3} $\\ 
$h	$ & $	0.705$ & $ 4.10\cdot 10^{-3}$ & $3.13\cdot 10^{-3} $ & $3.59\cdot 10^{-3}  $& $2.92\cdot 10^{-3}  $\\ 
$P_{s,5}	$ & $	1.07099$ & $2.98\cdot 10^{-2}$ & $2.68\cdot 10^{-2} $ & $2.77\cdot 10^{-2}  $& $ 2.60\cdot 10^{-2}$\\ 
$P_{s,6}	$ & $	1.04687$ & $2.89\cdot 10^{-2}$ & $2.11\cdot 10^{-2}  $ & $2.33\cdot 10^{-2}  $& $ 2.0\cdot 10^{-2}$\\ 
$P_{s,7}	$ & $1.02329$ & $2.00\cdot 10^{-2}$ & $1.73\cdot 10^{-2}  $ & $1.84\cdot 10^{-2}  $& $1.69\cdot 10^{-2} $\\ 
$P_{s,8}	$ & $	1.00024$ & $1.92\cdot 10^{-2}$ & $1.76\cdot 10^{-2}  $ & $ 1.86\cdot 10^{-2} $& $ 1.73\cdot 10^{-2}$\\ 
$P_{s,9}	$ & $	0.97771$ & $2.50\cdot 10^{-2}$ & $2.31\cdot 10^{-2}  $ & $ 2.43\cdot 10^{-2} $& $ 2.22\cdot 10^{-2}$\\ 
$f_{\rm NL}$ & $5$ & $12.4$ & $ 6.42$ & $ 5.23$& $4.46 $\\ 
$w	$ &$ 	-1$ & $3.23\cdot 10^{-2}$ & $ 2.46\cdot 10^{-2}$ & $3.99\cdot 10^{-2} $& $2.27\cdot 10^{-2}  $\\ 

\hline
\end{tabular}
\caption{As Table \ref{tab:nsk5} but including CMB priors.}
\label{tab:nsk5cmb}
\end{table*}

Table~\ref{tab:sk1} (\ref{tab:nsk1}) shows the $1\sigma$ marginalized errors for the case of a standard (\texttt{PCHIP}) PPS, for a fiducial value $f_{\rm NL}=20$  for each of the DESI tracers as well as the error from the combination of all of them, using the multi-tracer technique.  
Even if such a value of the $f_{\rm NL}$ parameter ($f_{\rm NL}=20$) is larger than the expected sensitivity from future probes, it is still allowed by current large scale structure bounds on primordial non-gaussianities.  
Notice that, for the standard power law PPS, the expected error on $f_{\rm NL}$ is $19.9$, $10.1$ and $8.56$ for LRGs, ELGs and QSOs respectively, while for the case of  the \texttt{PCHIP} parameterization, one obtains $\sigma(f_{\rm NL})=32.2$, $13.3$ and $12.6$ respectively. 
Therefore, there is a large increase in the error on the non-gaussianity parameter, which can reach the $60\%$ level. Concerning the remaining cosmological parameters, they are barely affected. 
In some cases, their error is even smaller than in the standard power-law scenario. 
This is indeed the case of the equation of state parameter $w$, or $\Omega_b h^2$ and $\Omega_c h^2$ (the errors on the latter two parameters are smaller than in the standard PPS approach only when exploiting either ELGs or QSOs tracers). 
The combination of the data from the three tracers exploiting the multi-tracer technique alleviates the problem with the error on $f_{\rm NL}$, as the increase in the value of $\sigma(f_{\rm NL})$  when relaxing the assumption of a simple power-law PPS is  around $40\%$, rather than $60\%$. 

\begin{figure*}
\begin{tabular}{c c}
\includegraphics[width=0.4\textwidth]{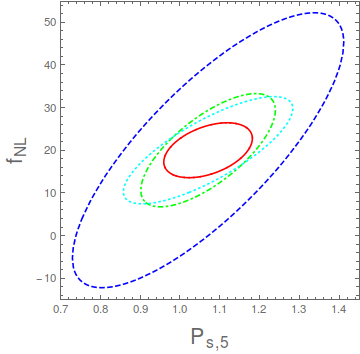}  &\includegraphics[width=0.4\textwidth]{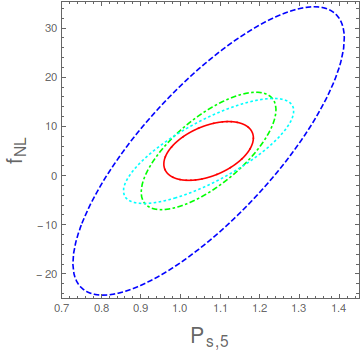}\\
\includegraphics[width=0.4\textwidth]{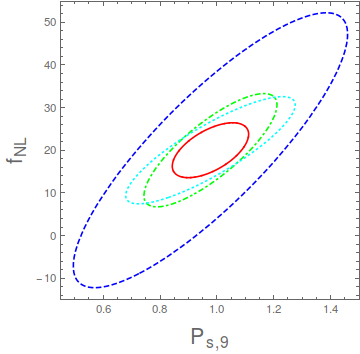}  &\includegraphics[width=0.4\textwidth]{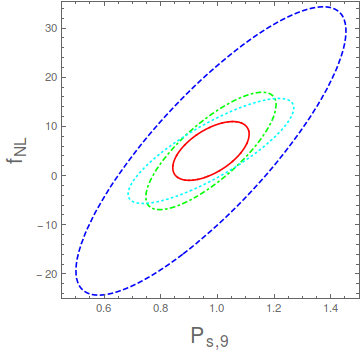}\\
\end{tabular}
\caption{The upper left (right) panel shows the $f_{\rm NL}-P_{s,5}$ degeneracy, for a fiducial cosmology with
$f_{\rm NL}=20$ ($f_{\rm NL}=5$), assuming $k_{\rm max}=0.1h$/Mpc. We show the $1-\sigma$ marginalized contours associated to the LRGs (in dashed blue lines),
ELGs (in dot-dashed green lines), QSOs (in dotted cyan lines) and multi-tracer (in solid red) Fisher matrix analyses. The bottom panels shows the analogous but in the ($f_{\rm NL}-P_{s,9}$) plane.}
\label{fig:deg}
\end{figure*}

The reason for this generic increase in the error of  $f_{\rm NL}$ is due to the the large degeneracies between the non-gaussianity $f_{\rm NL}$ parameter and the $P_{s,j}$ nodes, which get reduced when combining the tracers. 
The top and bottom panels of Fig.~\ref{fig:deg}  illustrate the large degeneracies between the non-gaussianity $f_{\rm NL}$ parameter, for the fiducial value $f_{\rm NL}=20$ and two of the $P_{s,j}$  nodes, $P_{s,5}$ and $P_{s,9}$.  
We only show here these two nodes, but similar degeneracies are obtained for the remaining nodes.

This degeneracy problem could a priori be solved in two ways, either exploiting smaller scales in the observed galaxy or quasar power spectra, or using CMB priors. 
In practice, going to the mildly non-linear regime would require new additional $P_{s,j}$  nodes and new degeneracies between these additional  $P_{s,j}$ nodes and the non-gaussianity parameter $f_{\rm NL}$ will appear. 
We have numerically checked that such a possibility  does indeed not solve the problem. 
Furthermore,  a non-linear  description of the matter power spectrum will depend on additional parameters, enlarging the number of degeneracies. 
In contrast, the CMB priors  on both the PPS parameters as well as on the dark matter and baryon mass-energy densities help enormously in solving the problem of the large degeneracies between the PPS parameterization and non-gaussianities. 
Tables~\ref{tab:sk1cmb} and \ref{tab:nsk1cmb} show the equivalent of \ref{tab:sk1} and \ref{tab:nsk1} but including CMB priors from the Planck mission 2013 data~\cite{Ade:2013zuv}. 
Notice that the impact of the Planck priors is largely more significant in the \texttt{PCHIP} parameterization case: 
the $f_{\rm NL}$ errors arising from the three different dark matter tracers when the CMB information is included are smaller in the \texttt{PCHIP} PSS description than in the standard power-law PSS modeling. 
When the multi-tracer technique is applied, the overall errors after considering Planck 2013 CMB constraints are very similar regardless on the PPS description and close to $\sigma(f_{\rm NL})\simeq 5$.

Table~\ref{tab:sk5} (\ref{tab:nsk5}) shows the $1\sigma$ marginalized errors for the case of a standard (\texttt{PCHIP}) PPS, for another possible non-gaussianity parameter fiducial value, $f_{\rm NL}=5$, from each of the DESI tracers, as well as the error arising from the combination of all of them using the multi-tracer technique.  As in the case of $f_{\rm NL}=20$, the error on the non-gaussianity parameter is increased,  
reaching in some cases a $60\%$ increment. The results are very similar to those obtained and illustrated  before for larger non-gaussianites. 
The errors on the other cosmological parameters remain unaffected under the choice of the PPS parameterization. The dark energy equation of state parameter is extracted with a smaller error in the \texttt{PCHIP} PPS case,  and also $\Omega_b h^2$ and $\Omega_c h^2$ are determined with a smaller error in that case while dealing with either ELGs or QSOs tracers. 
The multi-tracer technique provides a reduction on the $f_{\rm NL}$ error similar to that obtained in the $f_{\rm NL}=20$ case. 
The top and bottom right panels of Fig.~\ref{fig:deg} illustrate the large degeneracies between the non-gaussianity $f_{\rm NL}$ parameter and two of the $P_{s,j}$  nodes, $P_{s,5}$ and $P_{s,9}$ for the fiducial value $f_{\rm NL}=5$. 
Notice that the degeneracy pattern appears to be independent of the value of $f_{\rm NL}$. 
The addition of the CMB priors brings the errors on all the cosmological parameters ($f_{\rm NL}$ included) to the same values in both PPS parameterizations (standard power-law and  \texttt{PCHIP} PPS prescriptions), as shown in Tabs.~\ref{tab:sk5cmb} and \ref{tab:nsk5cmb}.  

We now perform an additional forecast.  We focus here on the shift induced in the local non-gaussianity  parameter $f_{\rm{NL}}$,
which we set to zero in the two cosmologies ${\cal
M}$ and ${\cal M}'$. For the purpose of this analysis, we fix
all of the $P_{s,j}$ to their best-fit values according to the Planck 2013 results, for the case of the ${\cal M}'$ cosmology.  A
shift in $f_{\rm{NL}}$ is expected to compensate for the fact that
the $P_{s,j}$ \texttt{PCHIP} nodes are additional parameters in
${\cal M}$, while not being considered as free parameters in the ${\cal M}'$
analysis.  Therefore, we displace the $P_{s,j}$ parameters (with $j=5,..,9$) from their fixed
fiducial values in $\cal M'$, i.e.  we are adding them as additional
parameters in the cosmological model (i.e. to be determined by
data). Referring to the notations of Eq.~(\ref{offset}), using a
shift $\delta\psi_{P_{s,j}}=0.1$, which is smaller than the
$1\sigma$ expected errors (see Tabs.~\ref{tab:nsk1} and
\ref{tab:nsk5}), we obtain that the corresponding shift in the
$f_{\rm NL}$ parameter is $\delta \theta_{f_{\rm NL}}\simeq 2.5$,
regardless of the exploited dark matter tracer. This is a quite
large displacement of the local non-gaussianity parameter which will
induce a non-negligible bias in reconstructing the inflationary
mechanism.  While the remaining cosmological parameters are also
slightly displaced with respect to their fiducial values, their shifts 
will not induce a misinterpretation of the underlying true cosmology.
The non-gaussianity  shift $\delta \theta_{f_{\rm NL}}$ could be a potential problem when
extracting the (true) value of the $f_{\rm NL}$ parameter not only
for the DESI survey, but also for future experiments with improved
sensitivities to non-gaussianities, such as
SPHEREx~\cite{Dore:2014cca}. The combination of all the three
possible DESI tracers leads to a smaller shift in the $f_{\rm NL}$
parameter ($\delta \theta_{f_{\rm NL}}\simeq 1.6$). If CMB priors
are applied, the shift is considerably reduced, $\delta
\theta_{f_{\rm NL}} \simeq 0.2$, which is close to the expectations
for non-gaussianities in the most economical inflationary models,
i.e. within single field slow-roll
inflation~\cite{Bartolo:2004if,Maldacena:2002vr}.

\section{Conclusions} 
\label{sec:concl}

While the simplest inflationary picture describes the power spectrum of the initial curvature perturbations $P_\mathcal{R}(k)$ by a simple power-law without features, there exists a large number of well-motivated inflation models that could give rise to a  non-standard PPS. 
The majority of these models will also generate non-gaussianities.  The large scale structure of the universe provides, together with the CMB  bispectrum,  a tool to test primordial non-gaussianites. 
Plenty of work has been devoted in the literature to forecast the expectations from upcoming galaxy surveys, such as the Dark Energy Instrument (DESI) experiment. 
The forecasted errors and bounds on the non-gaussianity local parameter $f_{\rm{NL}}$  are however usually derived under the assumption of a standard power-law PPS. 
Here we relax such an assumption and compute the expected sensitivity to $f_{\rm{NL}}$ from the DESI experiment assuming that both the precise shape of the primordial power spectrum and the non-gaussianity parameter need to be extracted simultaneously. 
If the analysis is restricted to large scale structure data, the standard errors computed assuming a featureless power spectrum are enlarged by $60\%$ within the \texttt{PCHIP} PPS parameterization explored here and when treating each of the possible dark matter tracers individually. 
Another potential problem in future galaxy surveys could be induced by the (wrong) assumption of a featureless PSS (while nature could have chosen a more complicated inflationary mechanism leading to a non-trivial PPS). 
If future data is fitted to the wrong PPS cosmology,  a shift in $|\delta \theta_{f_{\rm NL}}|\simeq 2.5$ would be inferred (for $k_{\rm max}=0.1 h$/Mpc) even if the true cosmology has $f_{\rm NL}=0$.  The multi-tracer technique helps in alleviating the former two problems. 
After combining all the DESI possible tracers, the forecasted errors on $f_{\rm{NL}}$ will be degraded by $40\%$ (when compared to the value obtained within the standard power-law PPS model) and the resulting shift will be reduced to  $|\delta \theta_{f_{\rm NL}}|\simeq 1.6$. 
The addition of Cosmic Microwave Background priors from the Planck 2013 data on the PPS parameters and on the dark matter and baryon mass-energy densities lead to a $f_{\rm{NL}}$ error which is independent of the PPS parameterization used in the analysis. 
After considering CMB priors, the value of the shift $|\delta \theta_{f_{\rm NL}}|$ is $0.2$, which is of the order of standard predictions for single-field slow-roll inflation~\cite{Bartolo:2004if,Maldacena:2002vr}.

\section{Acknowledgements}

The authors would like to thank Roland de Putter for very useful comments on the manuscript. 
This work was supported by the European Union Program FP7 ITN INVISIBLES (Marie Curie Actions, PITN-GA-2011-289442). 
O. M. is supported by Program PROMETEO II/2014/050, by the Spanish Grant No. FPA2011-29678 
and by the Centro de Excelencia Severo Ochoa Program, under Grant No. SEV-2014-0398, of the Spanish MINECO. 
L. L. H. was supported in part by FWO-Vlaanderen with the postdoctoral fellowship Project
No. 1271513 and the Project No. G020714N, by the Belgian Federal Science Policy Office 
through the Interuniversity Attraction Pole P7/37 
and by the Vrije Universiteit Brussel through the Strategic Research Program High-Energy Physics. 
The work of S. G. was supported by the Theoretical Astroparticle Physics research
Grant No. 2012CPPYP7 under the Program PRIN 2012 funded by the Ministero
dell'Istruzione, Universit\`a e della Ricerca (MIUR).

{}

\end{document}